\documentclass[a4paper,10pt]{article}

\begin{document}

\renewcommand{\thefootnote}{\fnsymbol{footnote}}

\begin{titlepage}
\begin{center}
\hfill LTH 865 \\
\vskip 10mm

{\Large
{\bf Extremal Black Holes, Attractor Equations, and Harmonic 
Functions}}

\vskip 10mm

\textbf{T. Mohaupt}\footnote{\textsf{Thomas.Mohaupt@liv.ac.uk}}
 and \textbf{K. Waite}\footnote{\textsf{Kirk.Waite@liverpool.ac.uk}}

\vskip 4mm

Theoretical Physics Division\\
Department of Mathematical Sciences\\
University of Liverpool\\
Liverpool L69 7ZL, UK \\

\end{center}

\vskip .2in 

\begin{center} {\bf ABSTRACT} \end{center}
\begin{quotation} \noindent
We review the construction of multi-centered black hole solutions 
through dimensional reduction over time. This method does not rely
on Killing spinor equations or gradient flow equations, but on
solving the second order field equations in terms of harmonic functions.
The black hole attractor equations are obtained directly from the field 
equations. 
\end{quotation}

\vfill

\end{titlepage}

\eject

\renewcommand{\thefootnote}{\arabic{footnote}}

\section{Extremal Black Holes}

The problem of constructing stationary solutions of a theory in 
$d+1$ dimensions can be reduced to the problem of constructing
solutions of a Euclidean theory in $d$ dimensions \cite{NKaBMG, Stelle}. 
If all relevant
fields fit into a scalar sigma model, one is essentially left with
solving the second order field equations for the scalars.
This method is very powerful
and allows the construction of, for example, multi-centered extremal black 
hole solutions in a very simple and systematic way, provided that the
scalar manifold satisfies certain integrability conditions. In \cite{MohWai}
this approach was applied to a class of five-dimensional Einstein-Maxwell-Scalar
type theories, which can be thought of as a natural generalization 
of (the bosonic sector of) five-dimensional vector supermultiplets.
In this article we review the construction of the solutions and some
of their properties.

\subsection{Dimensional Reduction}

Our starting part is a five-dimensional action of the following
form:
\[
S = \int d^5 x \sqrt{g_5} \left( \frac{1}{2} R_5 - \frac{3}{4}
a_{IJ}(\sigma) \partial_\mu h^I \partial^\mu h^J - \frac{1}{4}
a_{IJ}(h)  F^I_{\mu \nu} F^{J|\mu \nu} + \cdots \right)_{\hat{\cal V}(h)=1} \;.
\]
Besides gravity this theory contains $n$ gauge
fields $A^I_{\mu}$, with field strength $F^I_{\mu \nu}$, where
$I=1, \ldots, n$, and $n-1$ scalars. The scalars are assumed to take values on 
a smooth hypersurface of an $n$-dimensional manifold, which is given 
by the equation $\hat{\cal V}(h) = 1$, where $h^I$ are coordinates 
on the manifold, and 
where $\hat{\cal V}(h^I)$ is a homogeneous function of degree $p$.
The metric on the  $n$-dimensional space is of the 
form
\[
a_{IJ}(h) = - \frac{1}{p} \frac{\partial^2 \log \hat{\cal V}(h)}{
\partial h^I \partial h^J} 
\;.
\]
In the above action it is understood that the $n$ fields $h^I$ are 
subject to the hypersurface constraint $\hat{\cal V}(h)=1$. We could 
solve this constraint in terms of $n-1$ independent fields, but this
would turn out to be inconvenient. If $\hat{\cal V}(h)$ is a homogeneous
polynomial of degree $p=3$, the above terms are part
of the action of five-dimensional supergravity with
$n$ vector multiplets \cite{GST}. 
The bosonic part of the supergravity action also contains
a Chern-Simons term, but as we are only interested in stationary solutions 
carrying electric, but no magnetic charge, then the above 
truncation is consistent. We will allow that $p\not=3$ and thus 
consider a class of non-supersymmetric theories which have a similar 
structure, and which in particular are completeley determined by the 
choice of a prepotential $\hat{\cal V}(h)$. 

We reduce the five-dimensional theory with respect
to time, using the following decomposition of the metric
\[
ds_5^2 = - e^{2 \tilde{\sigma}}(dt + {\cal A}_m dx^m)^2 + e^{-\tilde{\sigma}}
ds_4^2 \;,
\]
and obtain the following four-dimensional Euclidean action:
\[ 
S = \int d^4 x \sqrt{g_4} \left( \frac{1}{2} R_4 - \frac{1}{2} 
N_{IJ}(\sigma) (\partial_m \sigma^I \partial^m \sigma^J 
- \partial_m b^I \partial^m b^J) + \cdots \right)\;.
\]
We have omitted the gauge fields because they correspond to magnetic
degrees of freedom of the five-dimensional theory. The $n$ scalars $b^I$
descend from the five-dimensional electro-static potentials $A_0^I$ and
have axionic shift symmetries $b^I \rightarrow b^I + C^I$, where $C^I$
are constant. The Kaluza Klein scalar $\tilde{\sigma}$ has been 
absorbed by rescaling the five-dimensional scalars. The resulting 
four-dimensional
scalars $\sigma^I = e^{\tilde{\sigma}} h^I$ are therefore $n$ independent
fields. The scalar metric $N_{IJ}(\sigma)$ is given by 
\[
N_{IJ} (\sigma) = - \frac{3}{2p} \frac{\partial^2 \log \hat{\cal V}(\sigma)}
{\partial \sigma^I \partial \sigma^J}  \;.
\] 
To solve the four-dimensional Euclidean equations of motion
we will assume that the metric is flat,
$g^{(4)}_{mn} = \delta_{mn}$. This implies that the five-dimensional
line element has a form which is known to occur for 
five-dimensional extremal black  holes. To solve the four-dimensional 
Einstein equations, we need to
impose that the energy-momentum tensor of the scalars vanishes identically.
This condition is easily seen to be equivalent to
\begin{equation}
\label{Constraint}
N_{IJ}(\sigma) \left( \partial_m \sigma^I \partial_n \sigma^J - 
\partial_m b^I \partial_n b^J \right) =  0 \;.
\end{equation}
The remaining scalar equations of motion take the form
\[
\partial^m ( N_{IJ} \partial_m \sigma^J) - \frac{1}{2} \partial_I N_{JK}
(\partial_m \sigma^I \partial^m \sigma^J - \partial_m b^I \partial^m b^J)
= 0 \;,\;\;\
\partial^m (N_{IJ} \partial_m b^J ) = 0 \;.
\]
The equations of motion of $b^I$ are the current conservation equations 
corresponding to the shift symmetries $b^I \rightarrow b^I + C^I$,
which are remnants of the five-dimensional gauge symmetries.
We observe that the ansatz
$\partial_m \sigma^I = \pm \partial_m b^I$ 
solves the constraint (\ref{Constraint}) and reduces the equations of 
motion to
\[
\partial^m ( N_{IJ} \partial_m \sigma^J )  =0  \;.
\]
It follows that if there exists dual fields $\sigma_I$, such that
$\partial_m \sigma_I = N_{IJ} \partial_m \sigma^J$, then the equations
of motion take the form of harmonic equations $\Delta \sigma_I =0$, 
and the solution can be expressed in terms of $n$ harmonic functions
$H_I (x)$. The existence of dual fields $\sigma_I$ requires that the
integrability condition $\partial_{[m} (N_{IJ} \partial_{n]} \sigma^J)=0$
is satisfied. This has two obvious classes of solutions: i) the
solution only depends one coordinate. This leads, for spherical symmetry,
to single-centered black holes solutions, or, for translational symmetry,
to domain-wall type solutions. ii) The scalar metric satisfies 
$\partial_{[I} N_{J]K}=0$, which is the integrability condition for the 
existence of a Hesse potential ${\cal V}$ for the metric 
$N_{IJ} = {\cal V}_{IJ} := \partial^2_{I,J} {\cal V}$. 
In this case no condition needs 
to be imposed on the space-time geometry. By construction, the 
scalar metric considered here have the Hesse potential: ${\cal V} = 
- \frac{3}{2p} \log \hat{\cal V}$. The dual scalars 
are given by the first derivatives of the Hesse potential:
\[
\sigma_I = {\cal V}_I \simeq \frac{\hat{\cal V}_I}{\hat{\cal V}} \;.
\]
Thus, given the Hesse potential, the dual fields can always be found
explicitly. However, it is not guaranteed that we can express
the original sclars $\sigma^I$ in closed form
in terms of the dual scalars $\sigma_I$, and, hence in terms of the 
harmonic functions. 
If we assume that the scalar fields
approach constant values at infinity, the simplest type of solution 
is given by multi-centered harmonic functions
\[
\sigma_I (x) = H_I(x) = h_I + \sum_{a=1}^N \frac{q_{aI}}{(x-x_{(a)})^2} \;.
\]
Whereas $h_I$ encode the values of the scalars at infinity, the
coefficients $q_{aI}$ measure the charges (with respect to the
shift symmetries) located at centers $x_{(a)}$. The total charges
$Q_I$ are obtained by summing over the centers. Note that the
charges associated with the shift symmetries can be written as 
surface integrals:
\begin{eqnarray}
Q_I &:=& \int d^4 x \partial^m ( N_{IJ} \partial_m b^J) =
\pm \oint d^3 \Sigma^m N_{IJ} \partial_m \sigma^J \nonumber \\
&=&
\pm \oint d^3 \Sigma^m \partial_m \sigma_I = 
\pm 2 \pi^2 \sum_{a=1}^N (-2) q_{aI} \;.  \nonumber
\end{eqnarray}

\subsection{Dimensional Lifting}

We now lift the solution of the four-dimensional Euclidean 
theory to five dimensions. The resulting line element is
\[
ds_5^2 = - e^{2\tilde{\sigma}(x)} dt^2 + e^{-\tilde{\sigma}} 
\delta_{mn} dx^m dx^n \;,
\]
where $\tilde{\sigma}$ is given in terms of the four-dimensional 
scalars by $e^{\tilde{\sigma}} = \hat{\cal V}(\sigma)^{1/p}$.
Rewriting $\sigma_I(x)  = {\cal V}_I = H_I (x)$ in terms of 
the five-dimensional scalars $h^I = e^{-\tilde{\sigma}} \sigma^I$, we obtain 
by 
\[
e^{-\tilde{\sigma}} \frac{\partial \hat{\cal V}}{\partial h^I } =
H_I \;,
\]
which are, in the supersymmetric case, precisely the so-called generalized
stabilization equations \cite{ChaSab}. 
If we approach the center at $x_{(a)}$, the
asymptotic behaviour of the harmonic functions is 
$H_I \approx\frac{q_{aI}}{r^2}$,
where $r = x - x_{(a)}$.
In the limit $r\rightarrow 0$ we obtain 
the five-dimensional stabilization or attractor equations \cite{ChaSab}
\[
Z_{a} \left. \frac{\partial \hat{\cal V}}{\partial h^I} 
\right|_{x=x_{(a)}} = q_{aI} \;,
\]
where $Z_{a} = \lim_{(x-x_{(a)}) \rightarrow 0} (r^2 e^{-\tilde{\sigma}})$. 
Thus the asymptotic solution is completely determined by the charges at the
center, which is a manifestation of the black hole attractor
mechanism \cite{Attractor}. Note that the near center
limit is (generically) finite, because $\sigma_I \sim r^{-2}$
implies, by homogeneity of $\hat{\cal V}$, that $e^{-\tilde{\sigma}} \sim 
r^{-2}$. Homogeneity can be used to solve for $Z_a$,
\[
Z_a = \frac{1}{d} q_{aI}h^I_{x=x_{(a)}} 
\]
The asymptotic value of $\tilde{\sigma}$ is given by 
$e^{-\tilde{\sigma}} \approx \frac{Z_a}{r^2}$, and the resulting 
asympototic 
metric at the center is $AdS^2 \times S^3$:
\[
ds^2 = - \frac{r^4}{Z_a^2} dt^2 + \frac{Z_a}{r^2} dr^2 + Z_a d \Omega_3^2
\]
This shows that we obtain, if all $Z_a$ are positive,  
a static configuration of 
extremal black holes with charges $q_{aI}$ and entropies
\[
S_a = \frac{1}{4}A_a = \frac{\pi^2}{2} Z_a^{3/2}  \;.
\]
Positivity of the  $Z_a$ imposes inequalities on the parameters
$q_{aI}$, which are, up to normalization and overall sign, the
electric charges carried by the centers. This is clear because 
the five-dimensional electric charges are proportional to the
charges associated to the four-dimensional shift symmetry.
The charges determine the asymptotic values of the scalars, and positive
$Z_a$ guarantee that the fixed point values approached by the
scalars at the center $x_{(a)}$ correspond to a 
regular point of the scalar manifold. Vanishing
$Z_a$ can be arranged by non-generic choices of the charges, for example
by putting sufficiently many charges to zero. This corresponds to 
a degenerate black hole horizon with vanishing area, and to scalar
fields running off to infinity on the scalar manifold.
For negative $Z_a$ this running off already occurs before 
the center is reached, resulting in a singularity of the space-time
metric.

The mass of the multi-centered black hole solution can be computed
by the ADM formula, with the result
\[
M_{ADM} 
= |h^I_\infty Q_I| \;,
\]
where $h^I_\infty$ are the values of the scalars at infinity, and 
where $Q_I$ are the total charges obtained by summing over the centers .  

\subsection{Discussion and concluding remarks}

We have seen that using dimensional reduction over time,
it becomes surprisingly simple to construct 
multi-centered extremal black hole solutions and to obtain 
the attractor equations which govern the near horizon fixed 
point behaviour. Our method does not require supersymmetry,
or the reduction of the field equations to first order equations,
but uses an integrability condition imposed on the scalar metric
in order to manipulate the second order field equations until
they have been reduced to harmonic equations. Note that the integrability
condition, while satisfied in supersymmetric theories, does not 
require that the underlying theory is supersymmetric, but applies
to a larger class. A deeper understanding why the method, which we
presented in a simple, pedestrian fashion in this article, works
so surprisingly well, can be obtained by a detailed analysis of
the geometry of the scalar sigma models occuring in the construction.
For this the reader is referred to \cite{EucIII,MohWai},
which contains a more comprehensive list of literature, which 
puts this work into the context of other recent work on 
black holes, and which discusses the four-dimensional 
Euclidean solutions in their own right.

{\bf Acknowledgement:}
T.M. would like to thank G. Zoupanos and all the other organizers
of the Corfu 2009 School and Workshop for the organization of a 
stimulating meeting and  the opportunity to present
this work.


\begin{thebibliography}{10}
\bibitem{NKaBMG} G. Neugebauer and D. Kramer, Ann der Physik (Leipzig)
24 (1969) 62. P. Breitenlohner, D. Maison and G. Gibbons, Comm. Math.
Phys. 120 (1988) 253.
\bibitem{Stelle} K. Stelle, {\em BPS Branes in Supergravity}, hep-th/9803116.
\bibitem{MohWai} T. Mohaupt and K. Waite, JHEP 0910 (2009) 058, arXiv:0906.3451.
\bibitem{GST} M. Gunaydin, G. Sierra and P.K. Townsend, Nucl. Phys. B 242
(1984) 244.
\bibitem{ChaSab} W.A. Sabra, Mod. Phys. Lett. A 13 (1998) 239, hep-th/9708103,
A.H. Chamseddine and W.A. Sabra, Phys. Lett. B 426 (1998) 36, hep-th/9801161.
\bibitem{Attractor} S. Ferrara, R. Kallosh and A. Strominger, Phys. Rev. 
D 52 (1995) 5412, hep-th/9508072. S. Ferrara, G.W. Gibbons and R. Kallosh, 
Nucl. Phys. B 500 (1997) 75, hep-th/9702103.
\bibitem{EucIII} V. Cort\'es and T. Mohaupt, JHEP 0907 (2009) 066, 
arXiv:0905.2844.
\end{thebibliography}
\end{document}